\documentclass[prd,onecolumn,preprintnumbers,nofootinbib]{revtex4}
\usepackage{graphicx}

\usepackage{color}
\usepackage{float}
\newcommand{\rthis}[1]{\textcolor{black}{#1}}
\usepackage{amsfonts,amsmath,amssymb}
\usepackage[plainpages=false, colorlinks=true, anchorcolor=blue, linkcolor=blue, citecolor=blue, bookmarks=false]{hyperref}
\usepackage{natbib}
\usepackage{enumitem}
\usepackage{lipsum}
\usepackage{multirow}
\usepackage{graphicx}
\usepackage{array}
\pdfoutput=1
\begin{document}
\newcommand{\bthis}[1]{\textcolor{black}{#1}}
\newcommand{\apjl}{Astrophys. J. Lett.}
\newcommand{\apjs}{Astrophys. J. Suppl. Ser.}
\newcommand{\aap}{Astron. \& Astrophys.}
\newcommand{\nar}{New  Astronomy Reviews}

\newcommand{\aj}{Astron. J.}
\newcommand{\araa}{Ann. Rev. Astron. Astrophys. } 
\newcommand{\mnras}{Mon. Not. R. Astron. Soc.}
\newcommand{\ssr}{Space Science Revs.}
\newcommand{\apss}{Astrophysics \& Space Sciences}
\newcommand{\jcap}{JCAP}
\newcommand{\pasj}{PASJ}
\newcommand{\pasp}{PASP}
\newcommand{\pasa}{Pub. Astro. Soc. Aust.}
\newcommand{\physrep}{Phys. Rep.}
\renewcommand{\arraystretch}{2.5}
\title{Low redshift calibration of the Amati relation using galaxy clusters}
\author{Gowri \surname{Govindaraj}}\altaffiliation{E-mail:ep20btech11007@iith.ac.in}

\author{Shantanu  \surname{Desai}}  
\altaffiliation{E-mail: shntn05@gmail.com}

\begin{abstract}
In this work, we use the angular diameter distances of 38 galaxy clusters with joint X-ray/SZE observation to circumvent the circularity problem in the Amati relation for Gamma-ray Bursts (GRBs). Assuming the validity of cosmic-distance duality relation, we obtain the luminosity distance from the  cluster angular diameter distance and use that to calculate the isotropic equivalent energy of two different GRB datasets, after restricting the GRB redshift range to $z<0.9$.
We then use these GRB datasets to test the Amati relation at the low redshifts using the galaxy cluster distances.  Our best-fit  Amati relation parameters   are consistent with a previous estimate for the same dataset. The intrinsic scatter which we  obtain for the two datasets is about 45\% and 15\%, and is comparable with that  found by other distance anchors used to study the Amati relation.
\end{abstract}

\affiliation{Department of Physics, Indian Institute of Technology, Hyderabad, Telangana-502284, India}
\maketitle

\section{Introduction}
Gamma-ray bursts (GRBs) are short-duration single-shot transients having  energies in the keV-GeV energy regime~\cite{Kumar,Luongo21}, which  were first discovered in the 1970s~\cite{firstgrb}. They are broadly classified into two categories:  short and long, depending on whether $T_{90}$ is less than or greater than two seconds~\cite{Kouv93}. Long-duration GRBs have been associated with core-collapse supernova~\cite{Bloom} and short GRBs with binary neutron star mergers~\cite{Nakar}.  However, there are still complex issues associated with the  classification and exceptions to the above dichotomy have also been found (see Refs.~\cite{Kulkarni,Bhave} and references therein).
GRBs are located at cosmological distances, with its maximum redshift  greater than nine~\cite{Levan}. However,  a distinct time dilation signature  in the GRB light curves (signature of cosmological expansion)  is yet to be unequivocally  demonstrated~\cite{Singh}.

For more than two decades, GRBs have also been proposed as standard candles using bivariate as well fundamental plane based correlations between different GRB observables in both the prompt as well the afterglow phase~\citep{Ito,Delvecchio,DainottiAmati,Moresco,Luongo21,Dainotti22,PradyumnaGRB}.
One of the most widely studied relations among these is the Amati relation, which posits a tight relation between the spectral
peak energy in the GRB  rest-frame ($E_p$) and the isotropic equivalent radiated energy ($E_{iso}$)~\cite{Amati02,Amati06}. An improved variant of the Amati relation has also been recently proposed using Gaussian Copula~\cite{Liu22}. 
However, given the paucity of GRBs at low redshifts, all the GRB correlation-based studies  can only be done  after assuming a cosmological model~\cite{Collazzi,Amati19}. Other systematics related to the Amati relation have been recently reviewed in Refs.~\cite{Moresco,Ito}.
To get around the above circularity problem in the Amati relation, two approaches have been used in literature. One way is to simultaneously constrain both the GRB correlation and cosmological model parameters~\cite{Amati08,Khadka20,Khadka21}. Alternately, 
a number of ancillary probes have also  been used to get model-independent estimates of distances corresponding to the GRB redshift,  such as Type 1a SN~\cite{LiangAmati,Kodama08,Demianski17,Liu22}, Cosmic chronometers~\cite{Montiel,Amati19,LuongoOHD,Luongo},  BAO $H(z)$ measurements~\cite{Luongo},  X-ray and UV luminosities of quasars~\cite{Dai21} and \rthis{also cosmography using BAO and Type 1a SN~\cite{Luongocosmo}}. In a similar vein, we use the angular diameter distances to galaxy clusters to calibrate the low redshift end of the Amati relation, without relying on an underlying cosmological model.

Galaxy clusters are the most massive virialized objects
in the universe~\cite{Allen2011,Borgani12,Vikhlininrev} and have proved to be  wonderful laboratories for
studying a wide range of topics from galaxy evolution to Cosmology to Fundamental Physics~\cite{Allen2011,Desai18,Boraalpha,BoraDesaiCDDR,Mendonca,PradyumnaRAR,Gopika,BoraDM,Chiu22}. However the redshift range of galaxy clusters is not as large as that of  GRBs with the redshift of the most distant galaxy cluster is currently less than  two. Therefore, we can only self-consistently test the validity of Amati relation at low redshifts ($z \leq 1$). Such a study could also be used  to probe the redshift evolution of the Amati relation~\cite{Dai21}. 

This manuscript is structured as follows.  We describe the GRB and galaxy cluster data used  for our analysis in Sect.~\ref{sec:data}. Our results are discussed in Sect.~\ref{sec:results}. We conclude in Sect.~\ref{sec:conclusions}.

\section{Datasets}
\label{sec:data}
\subsection{GRB datasets}
We carry out our analyses using two homogeneous GRB datasets compiled in literature. The first is the A220 dataset~\cite{Khadka21} and the other is the dataset compiled in Refs.~\cite{Demianski17,Demianski2}, which we refer to as the D17 dataset. The A220 dataset consists of 220 long GRBs (collated in Tables 7 and 8 in Ref.~\cite{Khadka21}). These span the redshift range $0.0331 \leq z  \leq 8.20$. The A220 dataset comes with the GRB redshift, peak energy in the rest frame ($E_p$), and bolometric fluence ($S_{bol}$).
The  D17  dataset,   consists of 162 long GRBs in the redshift range $0.125 \leq z \leq 9.3$. For each GRB, its redshift, distance modulus (using the closest Type 1a SN), $E_p$, and $E_{iso}$. The dataset is chosen based on joint detections by SWIFT/BAT and Fermi/GBM or Konus-WIND. Details of the selection criterion for selecting this dataset are  outlined in Ref.~\cite{Demianski17}.
Both A220 and D17 datasets also contain $1\sigma$ error bars for each of the aforementioned quantities. We also note that 24 GRBs are common among the two datasets.
Most recently, a study of the Amati relation for the low redshift end of the Amati relation was also done in another work~($0.125 \leq z \leq 1$)\cite{Dai21} (D21 hereafter).

\subsection{Galaxy cluster dataset}
The galaxy cluster dataset we use is the catalog of 38 clusters with joint Sunyaev-Zeldovich (SZ)~\cite{SZ} and X-ray observations compiled in the redshift range $0.14 \leq z \leq 0.89$~\cite{Bona06}. The SZ observations were carried out with  BIMA and OVRO. The X-ray observations were carried out using the Chandra X-ray observatory through the guaranteed time  program allocated to Leon van Speybroeck. More details of the X-ray and SZ observations and data reduction can be found in Refs.~\cite{Bona06,Bona04}. The angular diameter distances (and associated errors) were estimated by assuming a double-$\beta$ density profile~\cite{Mohr99} for the gas density distribution and spherical geometry for the cluster. This galaxy dataset has been widely used for a variety of model-independent cosmological tests such as cosmic distance-duality relation (CDDR)~\cite{Holanda10, Holanda11,Liang13,Santos}, tests of $\Lambda$CDM vs $R_h=ct$~\cite{Melia}, constraints on dark energy~\cite{ChenRatra} etc. In this work, we use the angular diameter distances of these clusters along with CDDR to get the luminosity distance corresponding to the redshift of a particular GRB.

\section{Analysis and Results}
\label{sec:results}
 For the analysis done in this work, we rewrite the Amati relation as a linear regression relation using the logarithms of  $E_{iso}$ and $E_p$, similar to D21:\footnote{Note that in some other works~\cite{Khadka21}, the regression variables $x$ and $y$ are flipped compared to the convention used here.}:
 \begin{equation}
 y = a x  + b
 \label{eq:amati}
 \end{equation}
 where $y \equiv \log (E_p (keV))$ and 
 $x \equiv \log (E_{iso} (erg))$.
 Note that $E_p$ is related to the observer frame peak energy given by $E_p= E_p^{obs}(1+z)$. Also 
 $E_{iso}$ is related to the bolometric fluence $S_{bolo}$  according to: 
 \begin{equation}
 E_{iso}=4 \pi d_L^2 S_{bol} (1+z)^{-1},
 \label{eq:eiso}
 \end{equation}
where $d_L$ is the luminosity distance corresponding to the redshift ($z$).

\begin{figure}
\centering
\includegraphics[width=0.5\textwidth]{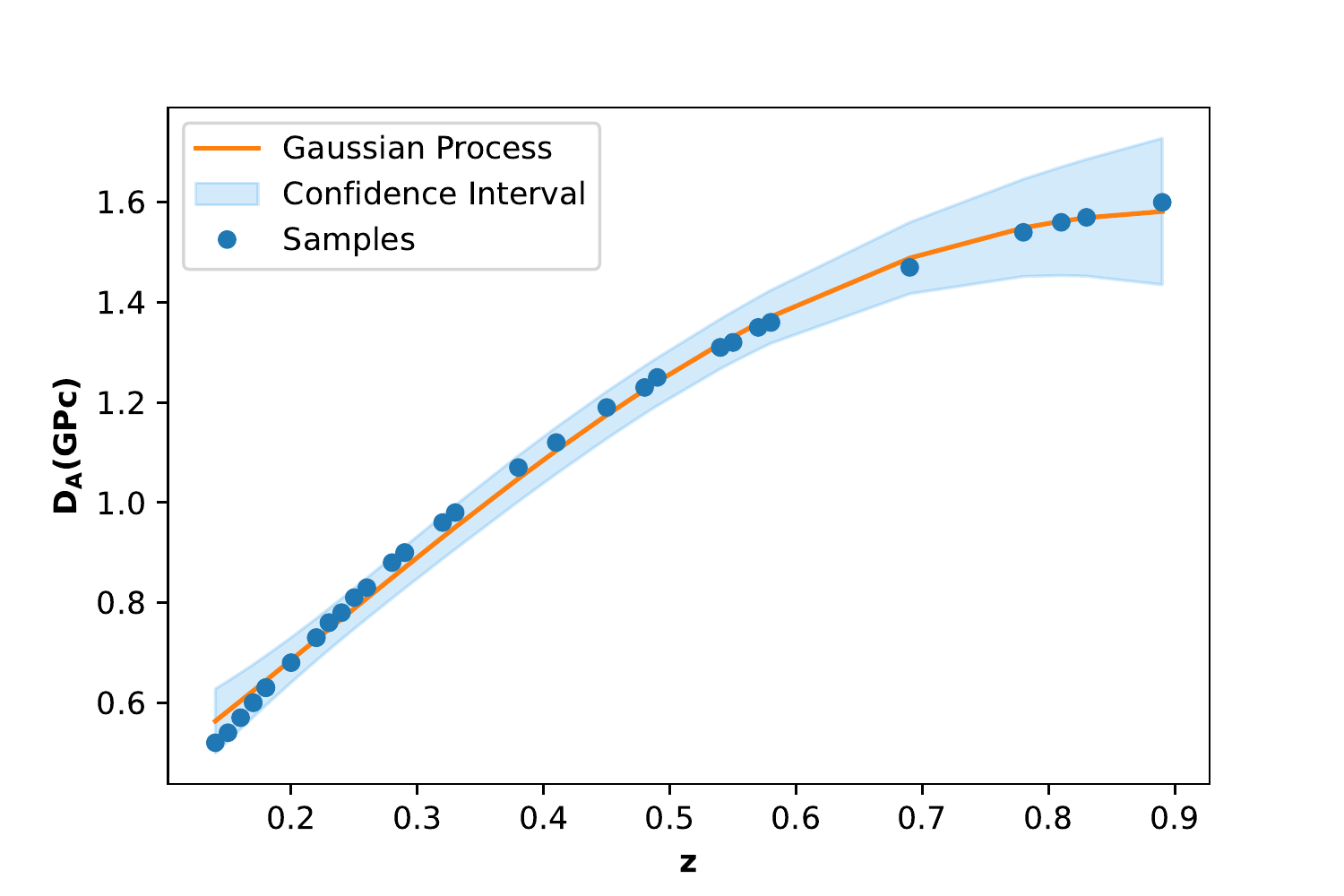}
\caption{Non-Parametric reconstruction of $D_A$  with GPR using data for   38 galaxy clusters with joint X-Ray/SZ observations~\cite{Bona06}. The GPR implementation was done using the {\tt HCGP} code~\cite{Camera}.}
\label{fig1}
\end{figure}

The first step in the analysis involves obtaining a model-independent estimate of $d_L$ from galaxy cluster observations. ~\citet{Bona06} provide the angular diameter distance ($D_a$) and redshift ($z$) data for 38 clusters  using joint X-ray/SZ observations.  With this data, we carry out a non-parametric reconstruction of $D_a$ as a function of $z$ using Gaussian Process Regression (GPR). GPR is a generalization of a Gaussian distribution, which is   characterized by a mean  and a covariance function (usually called the kernel function)~\cite{seikel12}. More details about GPR can be found in our previous works~\cite{HS,BoraDesaiCDDR,Borafg,BoraDM,Agrawal21,Mendonca,Mendonca2,Bora22}. For this  analysis,  we use the GPR code titled {\tt HCGP}~\cite{Camera} and employ the Radial Basis Function kernel. The GPR reconstruction of $D_A$ along with the associated  $1\sigma$ error bars can be found in Fig.~\ref{fig1}.  Although GPR can also be used for extrapolation, in this work we restrict our analyses to a subsets of the GRB datasets, which have the same redshift range as our cluster dataset so as to avoid any inaccuracy in extrapolation. This will also allow us to directly compare with the D21 results, who studied the evolution of the Amati relation by splitting the D17 dataset into four redshift subsamples.

Once we have reconstructed $D_a$ at any $z$, we can estimate $D_L$ from $D_A$ using the CDDR relation $D_L = D_A (1+z)^2$. The assumptions behind CDDR can be found in Ref.~\cite{BoraDesaiCDDR} and references therein. The CDDR has been validated   in a model-dependent fashion in a number of works~\cite{holanda19,BoraDesaiCDDR}.

For the D17 dataset, we recalculated  $E_{iso}$ from Eq.~\ref{eq:eiso} using the $D_L$ estimated from GPR and the distance modulus ($\mu$) provided in ~\cite{Demianski17}. For the A220 dataset, $E_{iso}$ was evaluated directly from Eq.~\ref{eq:eiso} using the $S_{bolo}$ provided for the  A220 dataset. Then using these values for $E_{iso}$ and the original $E_{p}$, we then  find the best-fit parameters of the Amati relation using Bayesian regression. For this purpose,
we use the following likelihood function based on Orthogonal Distance Regression,  which was previously  used  to analyze the Baryonic Faber-Jackson relation~\cite{Tian21} (See also ~\cite{Lelli19}) \footnote{Some previous works on testing the Amati relation ~\cite{Dai21,Moresco,Demianski17} have used a different likelihood (see for example, Eq. 11 in ~\cite{Dai21}). However, if one maximizes such a likelihood,  its parameters will diverge to infinity. This equation is usually attributed to ~\cite{Reichart}. However, we could not find this equation in ~\cite{Reichart}. Therefore, there must be some typographical error  in the expression for the likelihood in the above works.}

\begin{eqnarray}
-2\ln L &=& \large{\sum_i} \ln 2\pi\sigma_i^2 + \large{\sum_i} \frac{[y_i-(ax_i+b)]^2}{\sigma_i^2 (a^2+1)}
\label{eq:eq8}  \\
\sigma_i^2 &=& \frac{\sigma_{y_i}^2+a^2\sigma_{x_i}^2}{a^2+1}+\sigma_{s}^2
\end{eqnarray}
where $x$ and $y$ have the same  meaning as in Eq~\ref{eq:amati}; $\sigma_x$ and $\sigma_y$ denote the errors in $\log (E_{iso})$ and $\log (E_p)$, obtained using error propagation; and $\sigma_{s}$ denotes the loguniform intrinsic scatter which characterizes the tightness of the relation. We used uniform priors on $a$ and $b$, and log-normal priors on $\sigma_{s}$ : $a \in [0,1]$, $b \in [-30,-10]$, $\sigma_s \in [10^{-5}, 1]$. We sample the likelihood using the {\tt emcee} sampler~\cite{emcee}. 

\begin{figure*}
\centering
\includegraphics[width=\textwidth]{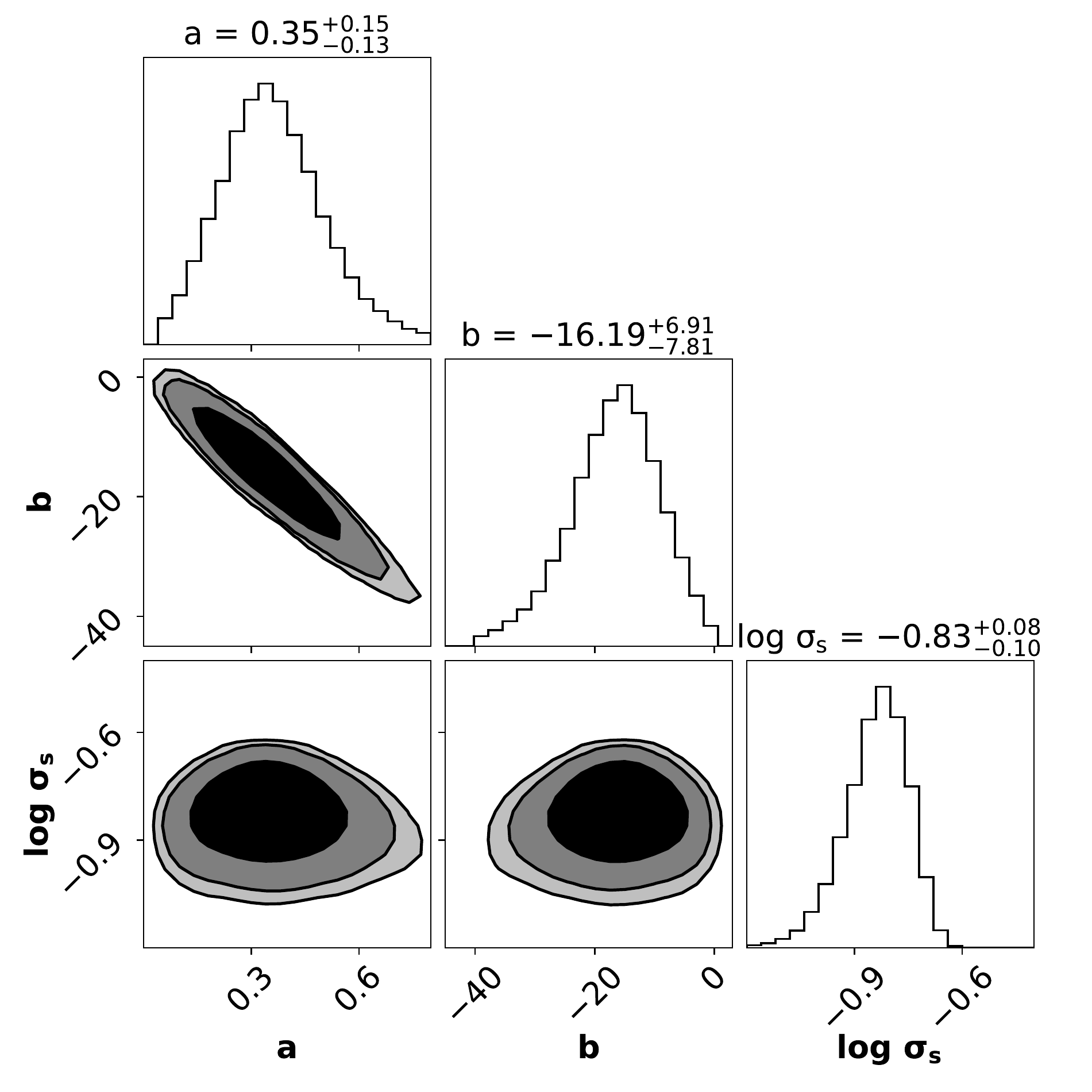}
\caption{Marginalized 68\%, 90\%, and 95\% credible intervals for  $a$, $b$, and $\ln \sigma_s$ (cf. Eq.~\ref{eq:amati}) for the  subset of D17  GRBs having $z<0.9$. The contours have been produced  using the {\tt Corner} package in Python. \rthis{Note that there is a  strong degeneracy between $a$ and $b$.}}
\label{fig2}
\end{figure*}

\begin{figure*}
\centering
\includegraphics[width=\textwidth]{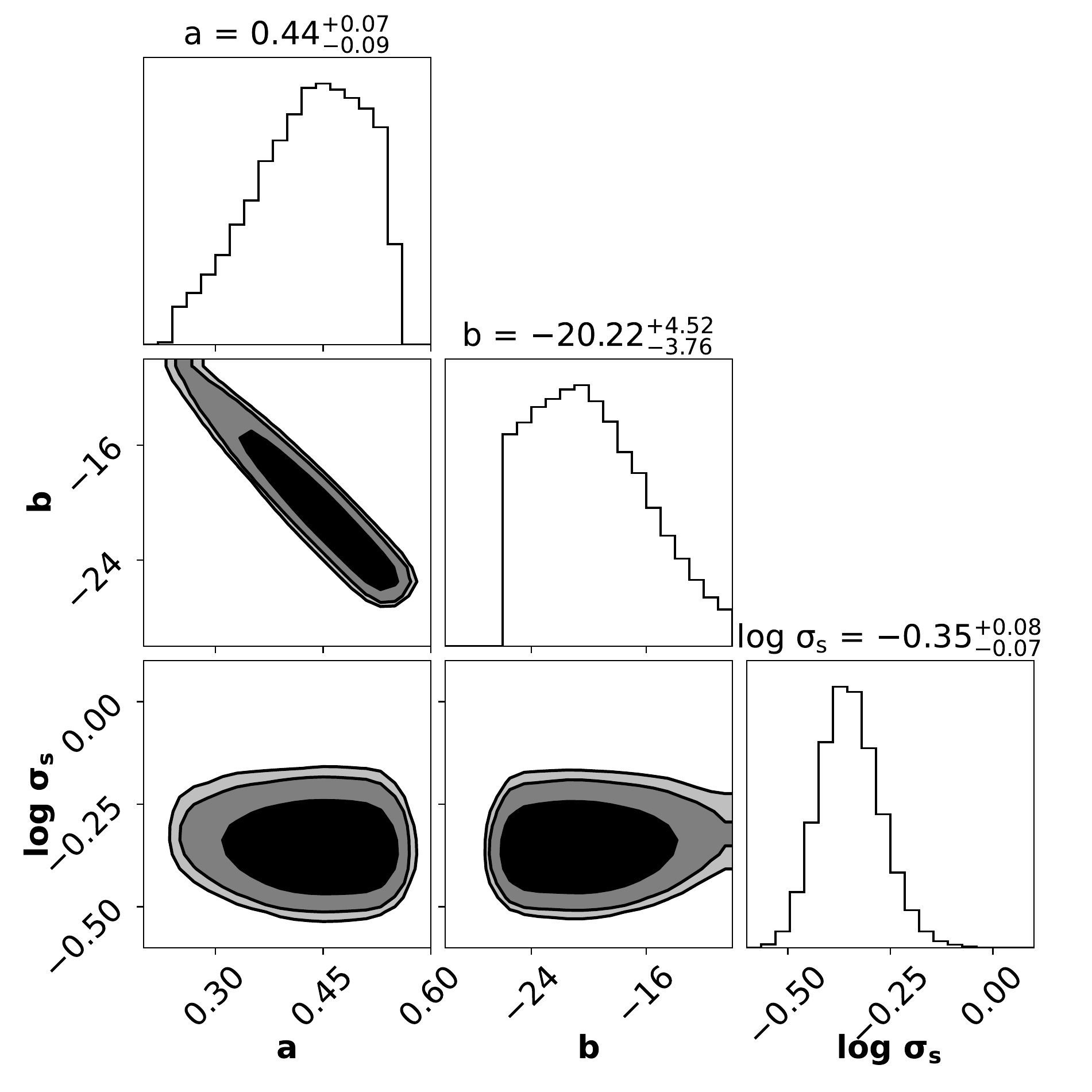}
\caption{Marginalized 68\%, 90\%, and 95\% credible intervals for  $a$, $b$, and $\ln \sigma_s$ (cf. Eq.~\ref{eq:amati}) for the subset of A220 GRBs having $z<1$. The contours have been produced  using the {\tt Corner} package in Python. \rthis{Similar to D17, there is again a  strong degeneracy between $a$ and $b$.} }
\label{fig3}
\end{figure*}

\begin{figure*}
\centering
\includegraphics[width=\textwidth]{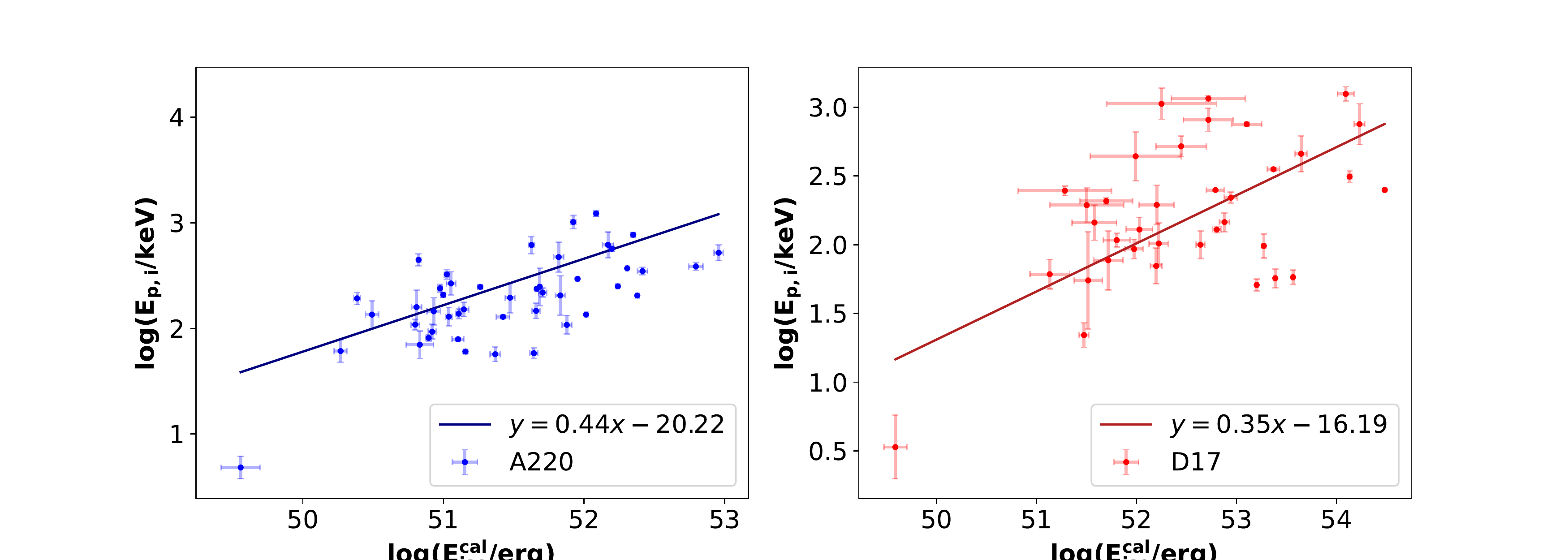}
\caption{Best-fit values for the Amati relation along with the data for both A220 and Di17 datasets having $z<1$.}
\label{fig4}
\end{figure*}

The marginalized 68\%, 90\%, and 95\% credible intervals can be found in Fig.~\ref{fig2} and Fig.~\ref{fig3} for the D17 and A220 GRB datasets with redshifts in the same range as the cluster catalog in ~\cite{Bona06}. The scatter plots for $E_p$ versus $E_{iso}$ using the updated values for $d_L$ for both the datasets are shown in Fig.~\ref{fig4}. A tabular summary of our results can be found in Table~\ref{table1}.

For the D17 dataset, the best-fit value of $a$ is equal to $0.35^{+0.15}_{-0.13}$ with an intrinsic scatter of 15\%. Our value of $a$ is consistent with the best-fit value found in D21 (for the lowest redshift sample) within $1\sigma$ (cf. Table 1 of D21). The corresponding scatter obtained by D21 (albeit with a different likelihood) for the D17 dataset using the lowest redshift sample is equal to 26\%. Therefore, our scatter is comparable to that obtained in D17.

For the A220 dataset, we find $a= 0.44^{+0.07}_{-0.09}$ with an intrinsic scatter of 45\%.  No other work has analyzed only the low redshift A220 subsample  and hence a direct comparison is not possible. However, ~\citet{Khadka21} find that the value of $a$ (in our notation) for the full A220 dataset (after doing a simultaneous fit to $\Lambda$CDM model) is given by $a=0.76\pm 0.044$, which is about $3.5\sigma$ discrepant with our analysis using the low-redshift subsample. The corresponding scatter obtained by ~\citet{Khadka21} for A220 dataset (using the full redshift sample), obtained from a  simultaneous fit to Cosmology and the Amati relation  is about 46\%. Most recently, a study of the Amati relation using A220 dataset was also carried out in ~\citet{Liu22} using the luminosity distance from Pantheon Type 1a SN, and the intrinsic scatter for the full A220 sample found to be about 50\%. Therefore, the scatter we obtained for this sample is comparable with the above estimates. Note however that the aforementioned works do not use the same likelihood as the one in this work.

Therefore, the intrinsic scatter for the Amati relation, which we find using the low redshift end of the A220 and D17 samples by using galaxy clusters as low redshift distance anchors is equal to 45\% and 15\%, respectively. This scatter is comparable to the same obtained using other model-independent probes as distances anchors to address the circularity problem in GRBs. However, given this somewhat large scatter, we conclude that the Amati relation is not as tight as some of the other fundamental plane GRB relations, when determined in a model-independent fashion. (See Table 2 of Ref.~\cite{PradyumnaGRB} and references therein for a comparison to  other GRB correlations.) Hence, it would be premature to use the Amati relation  as a stand-alone probe of precision Cosmology.

\begin{table}[h]

\begin{tabular}{ |c| c|c| c| }

\hline
\textbf{Dataset} & \textbf{a} &\textbf{b} & \textbf{Scatter} \\
\hline
A220 &$0.44^{+0.07}_{-0.09}$ &$-20.22^{+4.52}_{-3.76}$ & $0.45^{+0.091}_{-0.066}$ \\
D17 &  $0.35^{+0.15}_{-0.13}$ &$-16.19^{+6.91}_{-7.81}$ &$0.15^{+0.018}_{-0.030}$\\

\hline
\end{tabular}
\caption{\label{table1}Summary of our results for the Amati relation using  both the GRB datasets.}
\end{table}

\section{Conclusions}
\label{sec:conclusions}
In this work, we have used  galaxy cluster distances to address the circularity problem in the the Amati relation due to the paucity of GRBs at low redshifts. For this purpose, we use the $D_A$ of 38 galaxy clusters ($z<0.89$) obtained in a model-independent fashion using joint X-ray/SZ observations. From this, we obtain $D_L$ as a function of redshift, after positing the CDDR and using the GPR-based non-parametric regression technique. This $D_L$ was used to reconstruct $E_{iso}$ for two different GRB datasets (A220 and D17), using the low redshift GRB subset ($z<0.9$) in order to ensure that the redshift range overlaps with that of the galaxy cluster sample. Therefore, using galaxy clusters as anchors in this way, we can test the efficacy of the Amati relation at low redshift and probe its redshift evolution.

The best-fit credible intervals for the Amati relation parameters for the two GRB datasets used in this work, can be found in Fig~\ref{fig2} and Fig.~\ref{fig3}. We find that the best-fit parameters for one of the GRB datasets  are consistent with the results in  D21 within $1\sigma$, which also analyzed the same dataset, using  quasar UV and X-ray fluxes  as distance anchors instead of galaxy clusters.
 However, the intrinsic scatter which we obtain for both the datasets is quite large, viz. 44\% and 15\%. Although this scatter is comparable to that obtained using other cosmological probes as distance proxies or via doing a joint cosmology fit to the same datasets, its large value implies  that the Amati relation  cannot be used as a stand-alone precision probe of Cosmology. 
 
 \rthis{However, the Amati relation can still be used to test and constrain models of prompt emission of GRBs. For example the Amati relation can be reproduced from the internal shock model of prompt emission~\cite{Nava}  and conversely, this relation was also used  to constrain the dynamics of the flow and Lorentz factor~\cite{Nava}. A variety of studies have shown that the Amati relation can be explained based on the viewing angle in various jet models~\cite{Levinson}. This relation can also be explained within the alternative Cannonball model of GRBs~\cite{Dado}. A comprehensive review of constraints on various prompt emission models from the Amati relation can be found in ~\cite{Dai18,Ito}. The Amati  relation can also be used to discriminate between different GRB classes  and understand their nature and differences~\cite{Amati06,QinChen}.
 However, for all these studies  one must ensure that these are not affected by selection effects since the bursts with a high $E_{peak}$ and a relatively low $E_{iso}$ are more likely to remain undetected and be under-represented in observed samples~\cite{DainottiAmati}.}

On the GRB side, more breakthroughs should come from next generation missions such as SVOM~\cite{SVOM} and THESEUS~\cite{Theseus}. On the galaxy cluster side, one should expect a large number of additional X-ray/SZ observations from the next generation missions such as Athena~\cite{Athena} and Simons Observatory~\cite{Simons}. More precise tests of the Amati relation using GRBs in conjunction with galaxy clusters should therefore be possible within this decade.

\section*{Acknowledgments}
\rthis{We are grateful to the anonymous referee for useful comments and feedback on the manuscript.}
\bibliography{main}
\end{document}